# Visible light photonic integrated Brillouin laser

Nitesh Chauhan[1], Andrei Isichenko[1], Kaikai Liu[1], Jiawei Wang[1], Qiancheng Zhao[1], Ryan O. Behunin[2,3], Peter T. Rakich[4], Andrew M. Jayich[5], C. Fertig[6], C. W. Hoyt[6], Daniel J. Blumenthal[1*]

[1]Department of Electrical and Computer Engineering, University of California Santa Barbara, Santa Barbara, CA, USA.
[2]Department of Applied Physics and Materials Science, Northern Arizona University, Flagstaff, AZ, USA.
[3]Center for Materials Interfaces in Research and Applications, Northern Arizona University, Flagstaff, AZ, USA
[4]Department of Applied Physics, Yale University, New Haven, CT, USA
[5]Department Physics, University of California Santa Barbara, Santa Barbara, CA, USA.
[6]Honeywell International, Plymouth, MN, USA.
[*]Corresponding author (danb@ucsb.edu)

## ABSTRACT

Narrow linewidth visible light lasers are critical for atomic, molecular and optical (AMO) applications including atomic clocks, quantum computing, atomic and molecular spectroscopy, and sensing. Historically, such lasers are implemented at the tabletop scale, using semiconductor lasers stabilized to large optical reference cavities. Photonic integration of high spectral-purity visible light sources will enable experiments to increase in complexity and scale. Stimulated Brillouin scattering (SBS) is a promising approach to realize highly coherent on-chip visible light laser emission. While progress has been made on integrated SBS lasers at telecommunications wavelengths, barriers have existed to translate this performance to the visible, namely the realization of Brillouin-active waveguides in ultra-low optical loss photonics. We have overcome this barrier, demonstrating the first visible light photonic integrated SBS laser, which operates at 674 nm to address the $^{88}Sr^+$ optical clock transition. To guide the laser design, we use a combination of multi-physics simulation and Brillouin spectroscopy in a 2 meter spiral waveguide to identify the 25.110 GHz first order Stokes frequency shift and 290 MHz gain bandwidth. The laser is implemented in an 8.9 mm radius silicon nitride all-waveguide resonator with 1.09 dB per meter loss and Q of 55.4 Million. Lasing is demonstrated, with an on-chip 14.7 mW threshold, a 45% slope efficiency, and linewidth narrowing as the pump is increased from below threshold to 269 Hz. To illustrate the wavelength flexibility of this design, we also demonstrate lasing at 698 nm, the wavelength for the optical clock transition in neutral strontium. This demonstration of a waveguide-based, photonic integrated SBS laser that operates in the visible, and the reduced size and sensitivity to environmental disturbances, shows promise for diverse AMO applications.

# INTRODUCTION

Ultra-narrow linewidth (UNLW) visible light lasers provide the spectral purity required for precision atomic, molecular and optical (AMO) applications including atomic clocks[1,2], atomic and molecular spectroscopy[3–5], and quantum sensing[1,6,7]. Historically, it has been necessary to use macroscopic laser systems locked to large optical reference cavities to obtain the low phase noise and high frequency stability needed to address narrow optical clock transitions in atoms[8,9]. While providing state of the art performance[1,10,11], these lab-scale systems pose challenges for atomic and molecular experiments of ever-growing complexity, and for a portable or even autonomous optical clock. There is a need for lasers that are smaller and more reliable so that experiments can scale up, and in general, photonic integration will lead to reduced size, weight, and power consumption, as well as reduced sensitivity to environmental disturbances. Photonic integration is a promising approach to miniaturize laser systems as well as improve their reliability[12–14], thereby enabling systems with larger number of entangled atoms[15,16], higher sensitivity quantum sensors[6,17], higher precision positioning, timing and navigation[18], and probing of complex molecules[19–24].

Brillouin scattering (SBS) lasers, with their pump linewidth narrowing properties and ultra-low phase noise emission[25] are a promising candidate for AMO and quantum applications. Emission in visible has been achieved with fiber optic based resonators, exotic fiber, and bulk optic implementations[27-33]. Recently the coherence properties of an NIR fiber SBS laser were transferred to the visible to address the clock transition of strontium, however, this work required bulky, power inefficient, nonlinear frequency conversion[26]. To reduce system complexity and improve reliability, it is desirable to use a "direct-drive" approach, where the SBS laser directly emits at the desired visible wavelength, without intermediate conversion stages. Chip-scale SBS lasers operating in the NIR have exhibited impressive performance[34–41], achieving sub-Hz fundamental linewidth[34], 30 Hz integral linewidth over 100 ms, and $2\times10^{-13}$ fractional frequency stability[42]. To date, visible light emission in a photonic integrated SBS laser has remained out of reach. This lack of progress has been primarily due to barriers such as realizing ultra-low loss Brillouin-active planar waveguides in the visible. Overcoming these barriers, as well as realizing a visible light SBS laser in a wafer-scale integration platform, will reduce size and cost, as well as enable reliable scaling to more lasers for precision AMO science and applications.

We report the first demonstration, to the best of our knowledge, of visible light SBS lasing and Brillouin gain in a photonic integrated waveguide platform. This platform is compatible with foundry-level wafer-scale integration and can incorporate other photonic elements to realize systems on-chip[19]. The Brillouin gain medium is a high aspect ratio (20 nm tall by 2.3 µm wide) silicon nitride waveguide core surrounded by a thermal oxide on silicon lower- and tetraethoxysilane pre-cursor plasma-enhanced chemical vapor deposition (TEOS-PECVD) upper-silica cladding. The ability to realize visible Brillouin lasing using this waveguide structure leverages continuous SBS generation from long lifetime photons in an ultra-low loss (~1 dB/m) optical waveguide, an ultra-high quality factor (Q) resonator (>50 Million), and a short phonon lifetime due to the absence of acoustic waveguiding[34]. These properties lead to a sub-Hz fundamental laser linewidth and other benefits at 1550 nm[34]. We measure spontaneous Brillouin scattering in a 2 meter spiral waveguide, resulting in a 25.110 GHz Stokes frequency shift and 290 MHz Brillouin gain bandwidth. The shift and gain bandwidth are accurately predicted using a full-vectorial numerical simulation and enables accurate detection of the weak back-reflected spontaneous-Stokes pump heterodyne beat note. By combining our multi-physics simulations with measured data (Brillouin gain in fiber + waveguide, Brillouin gain in fiber), we can estimate (see Supplementary Information) the SBS gain to be 2.73 (W m)$^{-1}$. Brillouin lasing is demonstrated in an 8.9 mm radius silicon nitride bus-coupled ring resonator, with intrinsic quality factor (Q) = 55.4 Million and loaded Q = 27.7 Million[43] and a 3.587 GHz free spectral range (FSR) and a resonance-locked 674 nm semiconductor pump laser[31]. SBS lasing at 674 nm is verified by measuring a 14.7 mW on-chip pump threshold power and demonstrated linewidth reduction as the pump is increased from below to above threshold. The first order Stokes (S1) fundamental linewidth is determined by measuring the far-from-carrier frequency noise, with a resulting orders of magnitude linewidth decrease from the cold-resonator linewidth (i.e., cavity linewidth without light injected), confirming pump Brillouin phase noise suppression via long photon- and short phonon-lifetimes[25,34]. To highlight the versatility of this laser, we demonstrate 698 nm SBS lasing (the neutral strontium clock transition wavelength) using the same waveguide materials and design with mask-only changes to the ring diameter and change in the pump laser (698 nm).

To illustrate the utility of these integrated lasers, an integrated frequency stabilized SBS optical laser oscillator (OLO) can be used to "direct-drive" the strontium ion clock transition (see Fig. 1). A 674 nm pump laser is locked to the SBS resonator that in turn generates a backward propagation first order Stokes wave (S1) in the SBS resonator[14,44,45]. The Stokes wave is filtered by a three ring dual bus filter[46], and routed to an acousto-optic modular (AOM) that generates a single sideband (SSB), that in turn is locked to an on-chip optical reference cavity[47,48]. The associated electronics for laser tuning, SBS resonator locking, reference cavity stabilization, and AOM drive are illustrated as well as resonator tuning elements (e.g., thermal tuning). The stabilized 674 nm OLO beam is coupled to a $^{88}$Sr$^+$ ion in an electrostatic trap[49,50] using a waveguide to free-space grating coupler[51]. 3-dimensional (3D) cooling and repump beams are provided via on-chip or off-chip lasers and waveguide to free-space grating couplers. This scalable architecture could readily be to address an ion array by integrating an array of OLOs and Gratings on PIC.

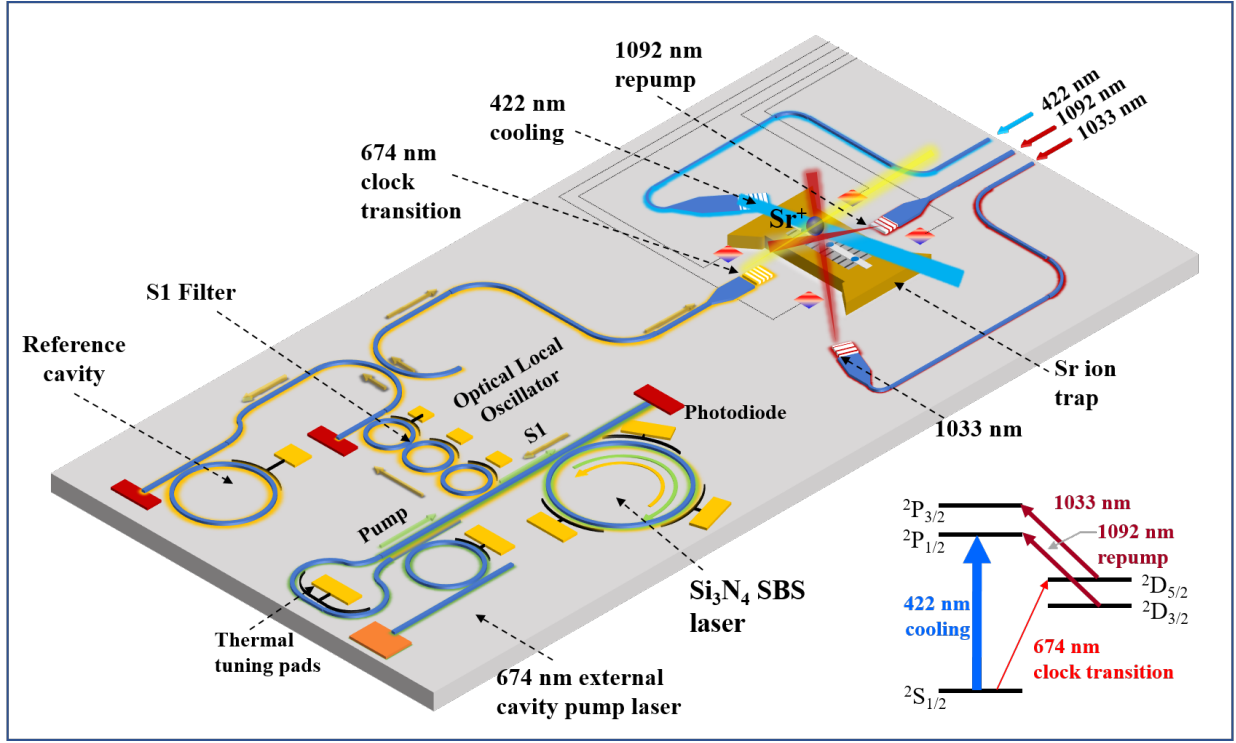

**Fig 1. Integrated optical clock with a stabilized SBS laser for addressing a trapped Sr ion.** Illustration of an integrated visible wavelength silicon nitride (Si₃N₄) waveguide stimulated Brillouin scattering (SBS) laser as a strontium ion (⁸⁸Sr⁺) clock transition optical laser oscillator (OLO) and cooling beam delivery interface. A heterogeneously integrated 674 nm external cavity Si₃N₄ tunable laser serves as the SBS laser pump. A three ring optical filter isolates and routes the SBS first order Stokes (S1) to a power splitter that sends a portion of S1 to an on-chip reference cavity. The reference cavity provides feedback to the pump laser and/or the SBS laser, to reduce the integral linewidth and stabilize the SBS laser for locking to the narrow atom transition. A single ⁸⁸Sr⁺ ion trap is shown in a chip-integrated trap. The ion is addressed with the 674 nm light, as well as other wavelengths necessary for clock operation. Fiber-coupled 1092 nm and 1033 nm lasers are converted to free-space beams from large-area grating emitters, incident on the ion, for state re-pumping[51–53].

## RESULTS

**Visible wavelength SBS laser resonator:** The SBS laser resonator is based on an ultra-low loss single mode Si₃N₄ waveguide that is designed to operate at 674 nm. The waveguide consists of a 20 nm tall and 2.3 μm silicon nitride core deposited and etched on a lower thermally grown oxide cladding on a silicon substrate, with a TEOS-PECVD deposited upper cladding[54] (Fig. 2a), for further details, see Methods, below. The spontaneous Brillouin gain frequency shift and Brillouin gain spectrum are measured using heterodyne detection between the pump and the backscattered signal in a 2 meter long spiral waveguide (Fig. 2b). Multi-physics vectorial simulations that incorporate actual materials and device parameters are used to predict the frequency offset and the Brillouin gain shape (red curve in Fig. 2c). The weak, back-scattered signal is measured by heterodyne detection of the pump-Brillouin beat note (Fig. 2c) with an electrical spectrum analyser (ESA), for details see the Supplementary Information. We measure a 25.110 GHz peak frequency shift and 290 MHz gain bandwidth, which agrees with our numerical simulations. The broad gain bandwidth is due to the continuous generation of photons in the ultra-low loss optical waveguide without acoustic waveguiding, which permits coupling to a continuum of bulk acoustic phonon states within the waveguide oxide cladding[34,37] with a resulting Brillouin gain coefficient of 2.73 (W m)[-1]. Brillouin scattering in the optical fiber used to deliver the 674 nm pump laser light is distinguished from the waveguide Brillouin scattering (Fig. 2c blue curve) by decoupling the fiber from the chip and making an independent measurement (blue curve of Fig. 2c), described further in the Supplementary Information).

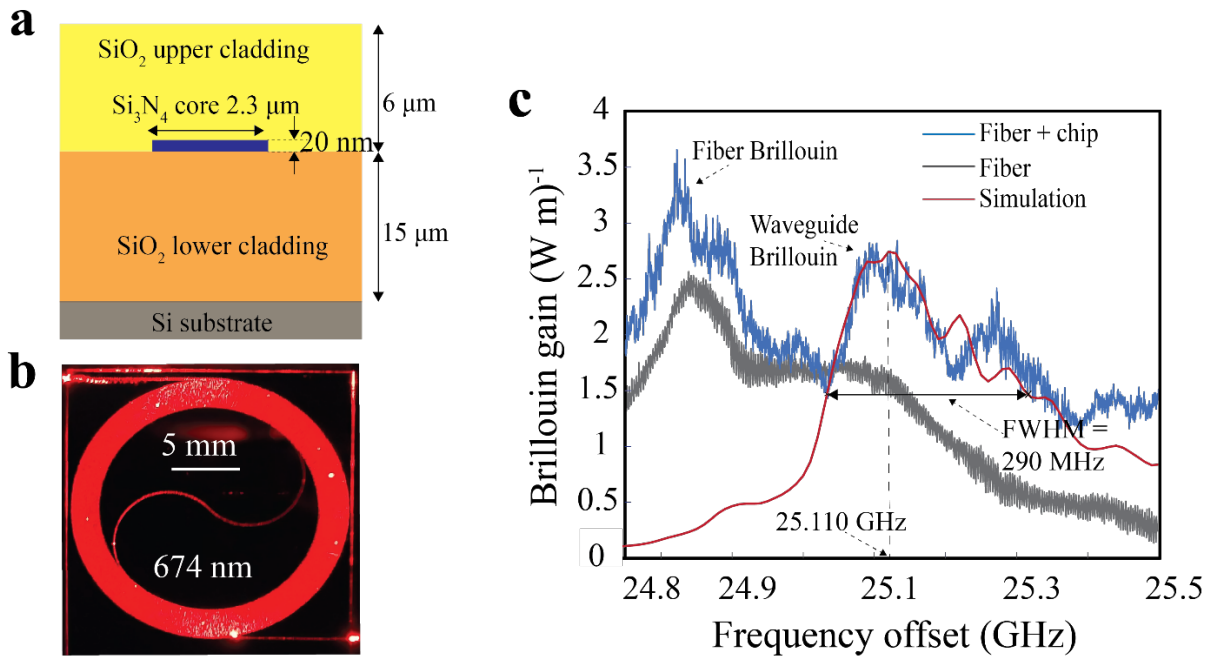

**Fig. 2. Waveguide design and 674 nm spontaneous Brillouin measurement: a,** Depiction of the waveguide cross section. **b,** Photograph of 2 meter single mode waveguide spiral used for the spontaneous Brillouin gain measurement, shown while illuminated with 674 nm light. 5 mm scale is shown for reference. **c,** Measured and simulated spontaneous Brillouin gain with 25.110 GHz first order Stokes (S1) frequency shift, 2.73 (W m)[-1] gain peak, and 290 MHz bandwidth. The measured blue curve shows the Brillouin contribution from both the fiber and silicon nitride waveguide, while the grey trace shows contribution from only the fiber, which confirms that the peak at 25.110 GHz is due to waveguide spontaneous Brillouin scattering.

The SBS laser resonator is a 8950.9 μm radius bus-coupled ring structure with a free-spectral range (FSR) designed to be 1/7 of the measured 25.110 GHz peak Stokes shift at 674 nm (Fig. 3b and 3c) and a bus-to-ring power coupling coefficient[34] $\kappa^2$ of ~1.5%. We design the ring to have multiple FSRs per Brillouin Stokes frequency shift, in order to reduce the fundamental linewidth through increased cavity volume, as well as to provide robustness to manufacturing variations[34]. An intrinsic Q of 55.4 Million and loaded Q = 27.7 Million at 674 nm is measured (Fig. 3d) using an RF calibrated MZI[54–56], yielding a propagation loss of 1.09 dB m[-1] (see Methods section).

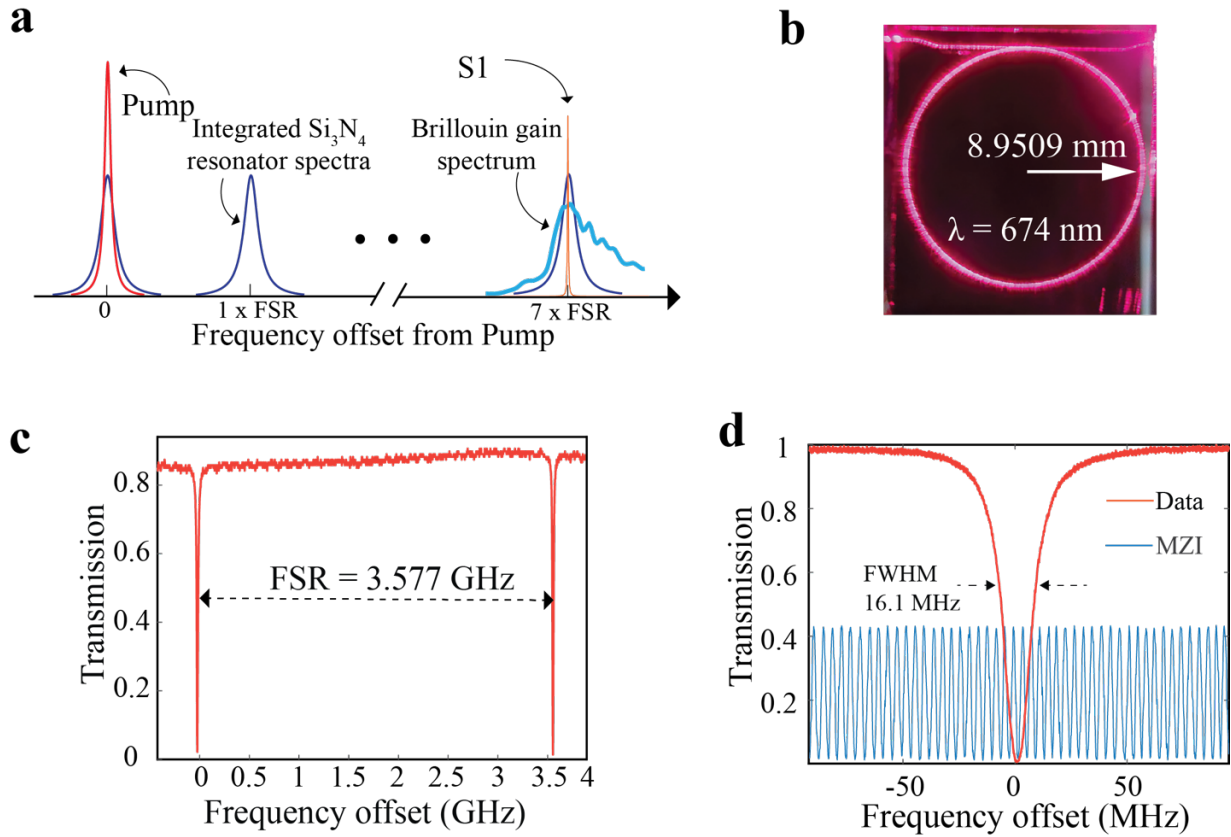

**Fig. 3. Stimulated Brillouin scattering (SBS) resonator design, FSR and quality factor**. **a**, Resonator design with free spectral range (FSR) = 3.587 GHz that is 7 times the 25.110 GHz spontaneous Brillouin peak gain shift. **b**, Photograph of silicon nitride waveguide resonator illuminated with fiber-coupled 674 nm laser light. The resonator radius is 8.9509 mm. **c**, Measured resonator FSR = 3.577 GHz at 674 nm. **d**, Quality factor (Q) measurement (red trace) performed using radio frequency (RF) calibrated unbalanced Mach-Zehnder interferometer (MZI) (blue trace). Full width half maximum (FWHM) linewidth = 16.1 MHz loaded Q = 27.7 Million and intrinsic Q = 55.4 Million and loaded Q = 27.7 Million at 674 nm.

**Visible light 674 nm SBS lasing:** The SBS laser resonator is pumped by a tapered amplifier (TA) that is seeded with 674 nm light from an external cavity diode laser (see experimental setup in Supplementary Information Fig. 3). The TA output is coupled to the waveguide SBS resonator through a high power fiber circulator. A maximum of 35 mW on-chip power is delivered to the waveguide bus, limited by the maximum 180 mW TA output power and ~4 dB fiber-to-facet coupling loss. The backward propagating SBS S1 signal is measured using a 3-port fiber optic recirculator located between the pump laser and the resonator input. The measured and simulated S1 powers are plotted versus the pump power in Fig. 4a. A clear S1 threshold is observed for an on-chip pump power of 14.7 mW where a 45% slope efficiency is measured, both in good agreement with our SBS model[55] (see Supplementary Information).

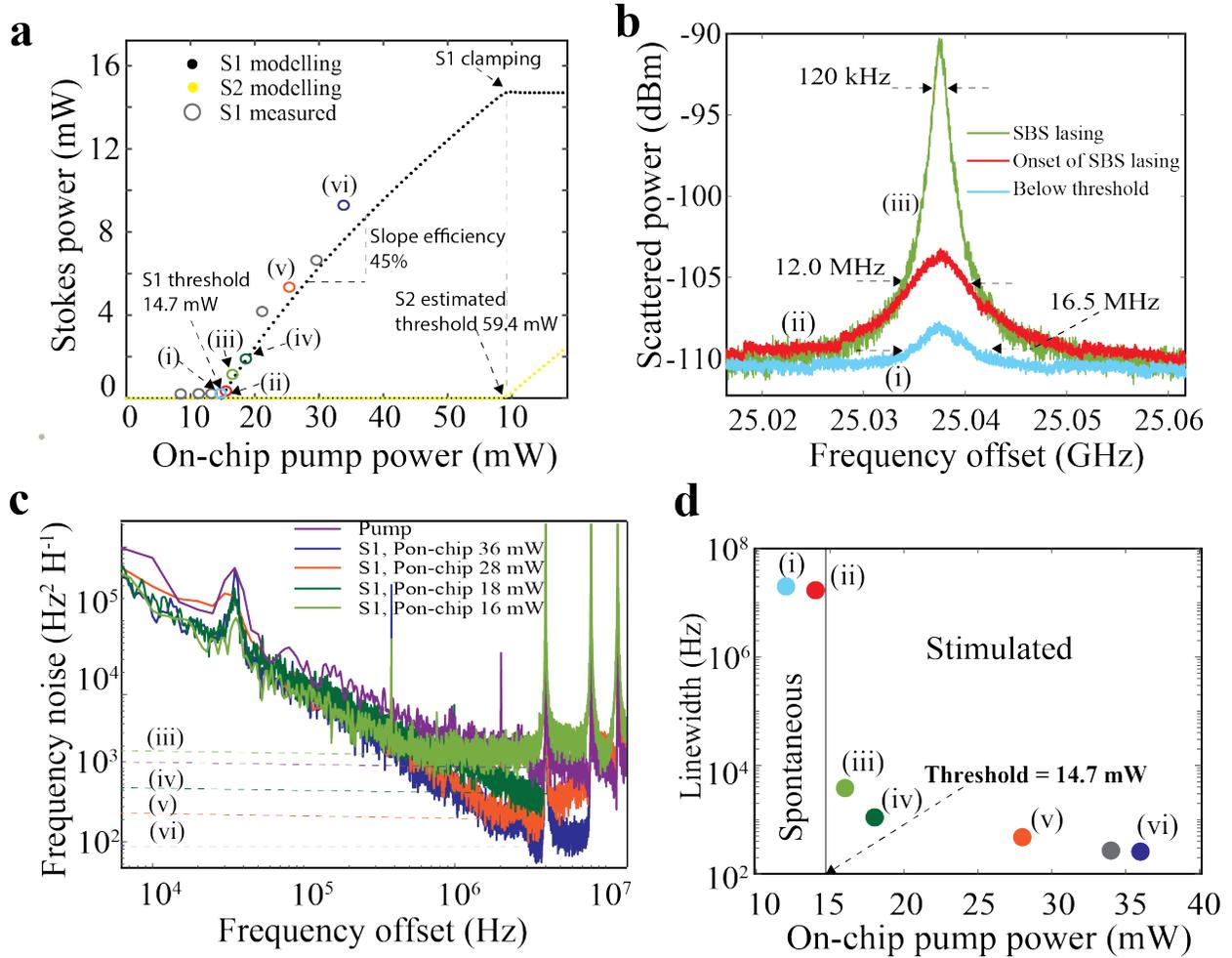

**Fig. 4. Stimulated Brillouin scattering (SBS) Stokes threshold, power and linewidth measurements. a**, Pump power on-chip ($P_{on-chip}$) vs. Stokes power. Measured first order Stokes (S1) laser threshold of 14.7 mW and 45% slope efficiency. Modelling of the S1 power (black dots) accurately predicts the measured S1 power as pump is increased from below to above threshold. The modelled S2 threshold is ~60 mW (yellow dots). **b**, Measurement of SBS first order Stokes (S1) emission linewidth, plotted on a logarithmic scale, below threshold (blue trace), just below threshold (red trace) and just after threshold (green trace). The Brillouin emission linewidth evolves from a spontaneous dominated linewidth of 16.5 MHz, which is approximately the SBS gain filtered by the cold-cavity resonator linewidth ~16 MHz, to the onset of stimulated Brillouin measuring a 12 MHz linewidth just below threshold, to a stimulated dominant 120 kHz linewidth just above threshold. **c**, Frequency noise measurements of S1 using a radio frequency (RF) calibrated fiber optic Mach-Zehnder interferometer (MZI) frequency discriminator. Fundamental linewidths are indicated by horizontal dotted lines that are tangent to the far-from-carrier frequency noise. Free-running pump frequency noise (purple trace) is for an unlocked pump (i.e., not locked to the stable cavity or SBS resonator). As the pump is increased, a decrease fundamental linewidth (curves ii to v) is measured. For this data, the back reflected pump is not optically filtered before frequency noise discrimination, leading to a contribution to the measured frequency noise of S1. Just above threshold, the conversion from pump to S1 is low, and the white noise floor at 16 mW pump (green) is as sum of pump and S1 and their beat note. As the on-chip pump, Pon-chip, is increased, the intra-cavity S1 photon number increases while the pump signal decreases to below 10dB of the Stokes for all other FN traces. **d**, Summary of beat note and fundamental linewidths from (b) and (c).

In addition to establishing the laser threshold, we demonstrate a decrease in the S1 emission linewidth as the pump power is increased from below threshold, through threshold, and above threshold[34,56]. Below threshold, the optical power spectrum is measured using a heterodyne beat note produced by mixing the backward propagating S1 with the pump on an electrical spectrum analyzer (ESA) (Supplementary Information). To minimize the contribution of the pump linewidth to the measured beat note we lock the pump laser to a commercial high finesse ultra-stable cavity (Stable Laser Systems). Well below threshold (i), the scattered light is produced by uncorrelated spontaneous scattering from thermal phonons, and is linearly filtered by the cavity resonance which is approximately 16.1 MHz. As threshold is approached, the spontaneous emission spectra, point (ii) in Fig. 4b, measures FWHM at 12.0 MHz, indicating the onset of stimulated emission, since the emission spectra is narrower than the cold-cavity resonance FWHM.

At just above threshold (iii), we see a dramatic narrowing of the linewidth to 120 kHz on the ESA as stimulated Brillouin scattering dominates the emission (Fig. 4b, trace (iii)), which is order 100x reduced from the passive cavity resonance linewidth. At all points above threshold, we measure the frequency noise of S1 using an optical frequency discriminator (OFD) (see Methods and Supplementary Information sections). The fundamental linewidth ($\Delta\nu$) is defined[34,38] as the far-from-carrier white frequency noise floor, in $Hz^2Hz^{-1}$, multiplied by $\pi$, where here the noise floor for each pump power input is indicated by horizontal dashed lines (iii) – (vi) in Fig. 4c. As the pump power increases beyond S1 threshold, the fundamental linewidth drops dramatically from 1.1 kHz (iv) to 269.7 Hz (vi). These linewidth results are summarized in Fig. 4d, indicating the integral linewidths for points (i) – (ii) below threshold, and the fundamental linewidths for the frequency noise curves in (iii) – (vi) in Fig. 4c. We were not able to provide the required on-chip pump power, 59.4 mW, to achieve lasing of the second order Stokes (S2). Future work will look further into noise properties measured using stabilized pump sources and exploring linewidth behavior as S1 approaches the S2 lasing threshold.

**Demonstration of 698 nm SBS lasing:** We have described a visible light integrated SBS laser approach and here describe briefly, demonstration of lasing at a second visible wavelength. The silicon nitride bandgap will support low loss for wavelengths longer than ~405 nm, making this laser a powerful tool to realize a broad range of atomic transition wavelengths by mask-only changes (the waveguide width, the ring diameter, the ring bus coupling gap), and the pump laser wavelength. To demonstrate this principal, we design and fabricate a 698 nm SBS resonator, a wavelength chosen to match the neutral strontium atom clock transition (Fig. 5a). The waveguide design and geometry are the same as for 674 nm (as verified by optical mode simulations, see Supplementary Information). Our multi-physics simulation predicts a 24.243 GHz Stokes shift and 300 MHz Brillouin gain spectra at 698 nm (Fig. 5b). Based on this expected shift we set the FSR such that the Stokes shift is separated from the pump by 7 FSRs (to maximize cavity volume and reduce sensitivity of alignment between Stokes shift and resonance), leading to a 9.4 mm radius resonator. We design a 3.4 μm bus to ring gap to operate in the under-coupled regime with a power coupling coefficient of ~1%. We measure a 12.7 MHz cavity resonance width, a 60 Million intrinsic Q and 33.8 Million loaded Q, and a 3.421 GHz FSR (see Fig. 5c and inset). First order SBS lasing is observed as S1 at the expected pump-Stokes frequency offset measured with an optical spectrum analyzer (OSA) as shown in Fig. 5c (the measured pump is reflected from the SBS resonator far facet). The S1-pump laser (Ti:sapphire) 23.892 GHz beat note is shown in Fig. 5c inset (see Supplementary Information for more details), and is 351 MHz off the simulated shift which can be explained by a slight offset between the peak of the gain and the narrow cavity resonance. The pump laser is free-running (i.e. stabilized neither to the resonator nor a supplementary Fabry-Perot optical cavity), so the beat note drifts at the 100 kHz level in tens of milliseconds. The on-chip pump power is 108 mW and the on-chip pump threshold power ($P_{th}$) is approximately 75 mW.

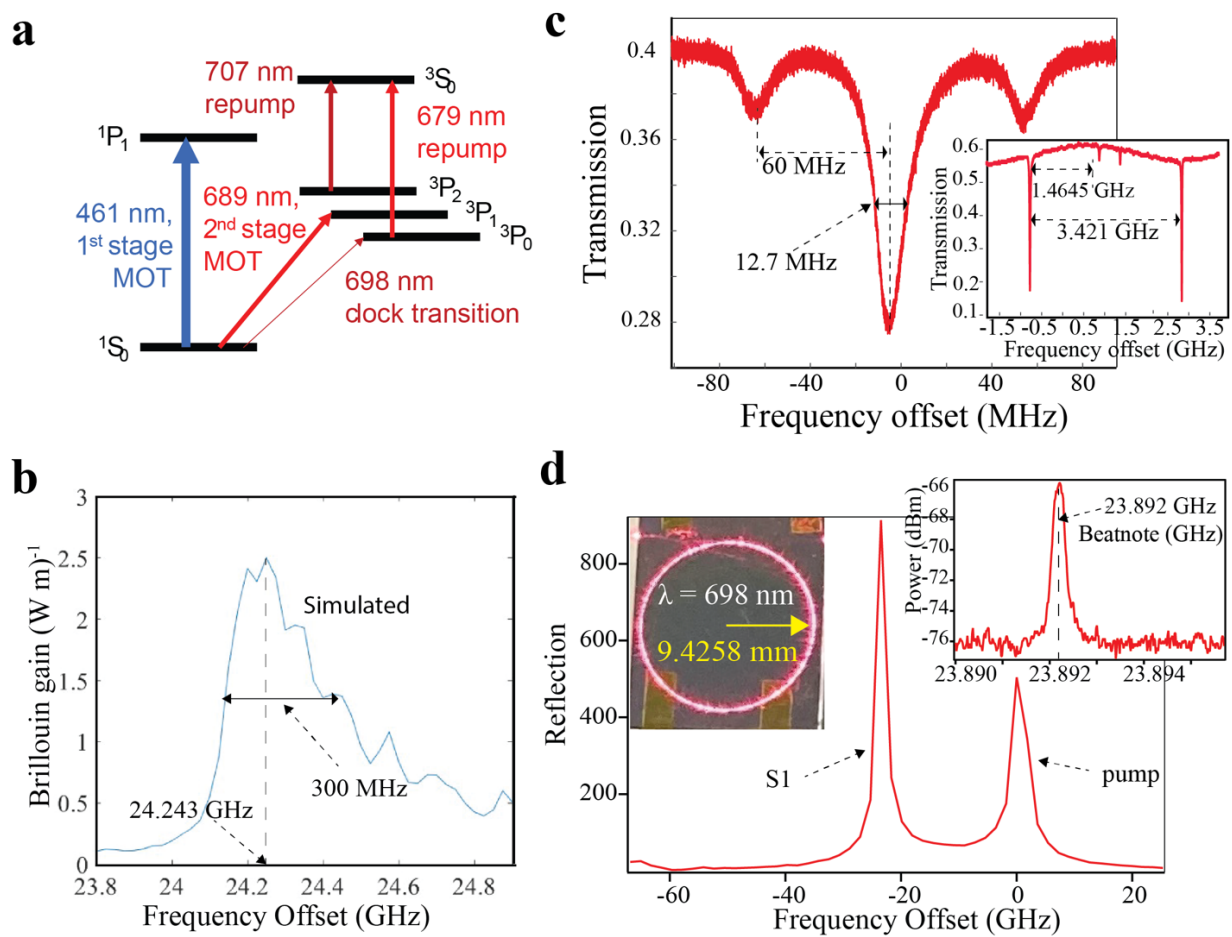

**Fig. 5. SBS lasing at 698 nm**. **a,** Neutral Sr transitions with 698 nm clock transition. **b,** Multi-physics simulation of waveguide spontaneous Brillouin gain spectra for pump at 698 nm. Brillouin gain spectrum width of ~300 MHz and frequency offset of 24.243 GHz. **c,** The measured full width half maximum (FWHM) of the resonator linewidth at 698 nm is 12.7 MHz and intrinsic Q of 60 Million and loaded Q of 33.8 Million. Inset shows the 3.421 GHz measured free spectral range (FSR) of the resonator for the transverse electric field (TE). **d,** First order Stokes (S1) and the pump measured on an optical spectrum analyzer (OSA), with inset showing the pump-S1 beat note measured at 23.892 GHz on an electric spectrum analyzer (ESA). Inset on left shows the 9.4 mm radius resonator fiber-coupled with 698 nm pump laser light during measurements.

# DISCUSSION

We have demonstrated the first, to the best of our knowledge, visible light SBS laser in a waveguide photonic integrated circuit. The 674 nm laser, based on an ultra-low loss silicon nitride bus-coupled ring resonator, is designed to serve as a "direct-drive", spectrally pure, chip-scale source that can couple to the $^{88}$Sr$^+$clock transition, without the need for intermediate frequency translation. To guide our design, we use a combination of multi-physics simulation and Brillouin spectroscopy to identify the first order Stokes (S1) frequency shift gain bandwidth, at 674 nm. Lasing is demonstrated with an on-chip 14.7 mW threshold, a 45% slope efficiency and Brillouin emission linewidth narrowing as the pump is increased from below threshold to above threshold, achieving 269.7 Hz fundamental linewidth at an on-chip pump power of 36 mW. Assuming the slope efficiency is constant and the calculated S2 threshold is 59 mW (Fig. 4a), at the onset of S2 lasing (threshold), we estimate it is possible to increase the S1 cavity photon number by a factor of 2x, resulting in an S1 fundamental linewidth reduction to ~153 Hz[55]. The measured frequency noise components result from various sources, including the SBS fundamental noise[38], intrinsic noise of the SBS cavity, pump amplitude noise that converts to SBS noise from locking the pump to the SBS resonator, and the technical noise sources in the pump laser and SBS resonator as well as pump amplitude to phase noise conversion from locking the laser to the cavity.

The ability to achieve SBS lasing in the visible comes about from the combination of long photon lifetime and ultra-high cavity Q using ultra-low loss optical waveguiding, and short phonon lifetime. This design enables lasing at other visible SBS emission wavelengths with mask-only changes and a change in the pump source. To illustrate this versatility, we demonstrate lasing at 698 nm, a wavelength suitable to probe long-lived transitions in neutral strontium. Looking forward, we will be characterizing the full frequency noise, coherence, and linewidths at other visible laser emission wavelengths.

In order to improve SBS laser efficiency and further reduce linewidth, our simulations show that with closer FSR matching to the Brillouin gain peak shift we can achieve a 7 mW threshold with the current design and generate higher S1 optical power. This can be achieved by adjustment of the SBS cavity using waveguide tuning technique including thermal[57] and piezoelectric[58]. Our simulations show that continued increase in S1 photon number up to S2 threshold will lead to a fundamental linewidth reduction to ~ 2 Hz. Other possible improvements to further reduce the linewidth include modulating the laser resonator with a grating, to split the second order Stokes (S2) resonance and prevent S2 emission and further increase in S1 optical power[55]. Given the transparency and bandgap of silicon nitride, and the low loss achievable down to ~405 nm, this platform will support a wide range of SBS photon-phonon interactions and as such, wavelengths for a variety atomic and molecular transitions. Future work will involve demonstrating this design across the broad range of silicon nitride waveguide transparency (e.g., Yb @ 578 nm, Ca + @ 729 nm and waveguides with higher bandgap that can support the UV (e.g., Al+ 267.4 nm).

## METHODS

**Fabrication process.** The substrate and lower cladding consist of a 15-μm-thick thermal oxide grown on a 100-mm diameter silicon wafer. The main waveguide layer is a 20-nm-thick stoichiometric $Si_3N_4$ film deposited on the lower cladding thermal oxide using low-pressure chemical vapor deposition (LPCVD). A standard deep ultraviolet (DUV) photoresist layer was spun and then patterned using a DUV stepper. The high-aspect-ratio waveguide core is formed by anisotropically dry etching the $Si_3N_4$ film in an inductively coupled plasma etcher using a $CHF_3/CF_4/O_2$ chemistry. After the etch, the wafer is cleaned using a standard Radio Corporation of America (RCA) cleaning process[59]. A 6-μm-thick silicon dioxide upper cladding layer was deposited in two 3-μm steps using plasma-enhanced chemical vapour deposition (PECVD) with tetraethoxysilane (TEOS) as a precursor, followed by a final two-step anneal at 1050 °C for 7 hours and 1150 °C for 2 hours.

**Resonator linewidth measurements.** *674 nm resonator:* We used the same cateye diode tunable pump laser (from mogLabs) for these measurements. A ~50 m unbalanced fibre-based radiofrequency (RF) calibrated Mach-Zehnder interferometer (MZI) was used to measure the Q[60]. To calibrate the MZI free spectral range (FSR), an RF electro-optic phase modulator (EOM) was used to create two sidebands. While scanning across a resonance, the two sidebands are used to calibrate the MZI FSR. The MZI FSR is measured to be 3.99±0.02 MHz. The MZI is acoustically isolated to minimise noise in the fringes. We simultaneously scan the laser through both the MZI and the resonator and the MZI FSR provides a RF calibrated frequency reference for calibrating the resonance linewidth. *698 nm resonator:* We use a Ti:Sapphire laser at 698 nm to measure the Q. To calibrate the frequency, we use two different phase modulators to add sidebands at 1.4645 GHz for FSR measurements and at 60 MHz for Q measurements.

**Frequency noise measurements.**

We measured the frequency noise and fundamental linewidth of our laser with an optical frequency discriminator (OFD) consisting of a fiber based unbalanced MZI (UMZI) and a balanced photodetector. The frequency noise of the laser, $S_f(\nu)$ in ($Hz^2Hz^{-1}$) is related to the power spectral density of the detector output $S_{out}(\nu)$ in ($V^2Hz^{-1}$) as:

$$S_f(\nu) = S_{out}(\nu) \left( \frac{\nu}{sin(\pi\nu\tau_D)V_{PP}} \right)^2$$

where $\tau_D$ is the optical delay of the UMZI, $\nu$ is the frequency offset, $V_{PP}$ is the peak-to-peak voltage of the detector output. The fundamental linewidth $\Delta\nu = \pi S_w$ is determined by the value $S_w$, that is tangent to the lowest point of the $S_f(\nu)$ far-from-carrier noise measurements at frequencies typically above 1 MHz.

The S1 power from the reflection port of the circulator is sent into the acoustically isolated, 50m fiber-delay UMZI with a FSR of 3.99 MHz. Both UMZI outputs are connected to the balanced photodetector (Thorlabs PDB450A) with a bandwidth of 150 MHz in order to reduce the impact of intensity variations in the detector output. The power spectral density $S_{out}(\nu)$ of the detector difference (RF) output was measured using a digital sampling oscilloscope (Keysight DSOX1204G with 200 MHz bandwidth). The RF output triggers the scope at the quadrature operating point of the UMZI and the power spectral density data is averaged over 16 traces with a Hann window is applied. The frequency noise is calculated using equation (1).

## DATA AVAILABILITY

The data that support the plots within this paper and other findings of this study are available from the corresponding author on reasonable request.

## ACKNOWLEDGEMENT

This work was supported by DARPA MTO APhI contract number FA9453-19-C-0030. The views, opinions, and/or findings expressed are those of the authors and should not be interpreted as representing the official views or policies of the Department of Defense or the U.S. Government. Andrei Isichenko acknowledges the support from the National Defense Science and Engineering Graduate (NDSEG) Fellowship Program. The authors would like to thank Mingyu Fan, Sean Buechele and Michael Straus from the Department of Physics, UCSB for their help aligning the 674 nm pump laser.

## AUTHOR CONTRIBUTIONS

N.C., R.O.B., P.T.R., C.W.H, A.M.J, and D.J.B. prepared the manuscript. R.O.B. performed the multi-physics simulations for spontaneous Brillouin scattering. N.C. and K.L. did the simulation for Stokes power vs on chip power and threshold. N.C. did the waveguide, spiral and the resonator design. J.W. and Q.Z. fabricated the devices. N.C performed spontaneous Brillouin scattering measurements at 674 nm and N.C. and A.I. did the SBS laser threshold, frequency noise and linewidth measurements at 674 nm. C.W.H., C.F. and N.C. performed the SBS laser measurements at 698 nm. D.J.B., P.T.R., R.O.B, A.M.J. C.W.H. and C.F. supervised and led the scientific collaboration.

## COMPETING INTERESTS

The authors declare no competing interest.

# Supplementary Information

# Visible light photonic integrated Brillouin laser

Nitesh Chauhan[1], Andrei Isichenko[1], Kaikai Liu[1], Jiawei Wang[1], Qiancheng Zhao[1], Ryan O. Behunin[2,3], Peter T. Rakich[4], Andrew M. Jayich[5], C. Fertig[6], C. W. Hoyt[6], Daniel J. Blumenthal[1*]

[1]Department of Electrical and Computer Engineering, University of California Santa Barbara, Santa Barbara, CA, USA.
[2]Department of Applied Physics and Materials Science, Northern Arizona University, Flagstaff, AZ, USA.
[3]Center for Materials Interfaces in Research and Applications, Northern Arizona University, Flagstaff, AZ, USA
[4]Department of Applied Physics, Yale University, New Haven, CT, USA
[5]Department Physics, University of California Santa Barbara, Santa Barbara, CA, USA.
[6]Honeywell International, Plymouth, MN, USA.

[*]Corresponding author (danb@ucsb.edu)

## Supplementary Note 1: Introduction

In this Supplementary Information, we reveal more details on the measurement setup for the spontaneous and stimulated Brillouin measurement; we discuss in detail the modelling of stokes power as the on chip power is increased above threshold.

## Supplementary Note 2: Spontaneous Brillouin scattering measurement setup

Spontaneous Brillouin scattering is scattering by uncorrelated incoherent photon-phonon gratings resulting in very weak backscattered signals. To reliably measure the spontaneous backscattering signal a 40 dB gain radio frequency (RF) amplifier is used to boost the photodiode (PD) signal. The resolution bandwidth (RBW) of the electrical spectrum analyzer is set to 100 Hz to increase the sensitivity. 10 traces are averaged on the ESA to reduce the noise. A background trace is taken with no input to PD and all settings of ESA kept the same and subtracted from the measurements to remove instrument background noise. Traces are taken with device coupled and that yields Brillouin scattered signal from waveguide (2 m spiral) as well as ~ 0.5 m fiber between circulator and device facet. Another measurement is taken with the device uncoupled to identify the peak from the fiber section while keeping the power constant. The fiber peak remains unchanged while the waveguide peak disappears. The measured data is used to scale the multi physics simulation and gain is then extracted from the fitted simulated curve.

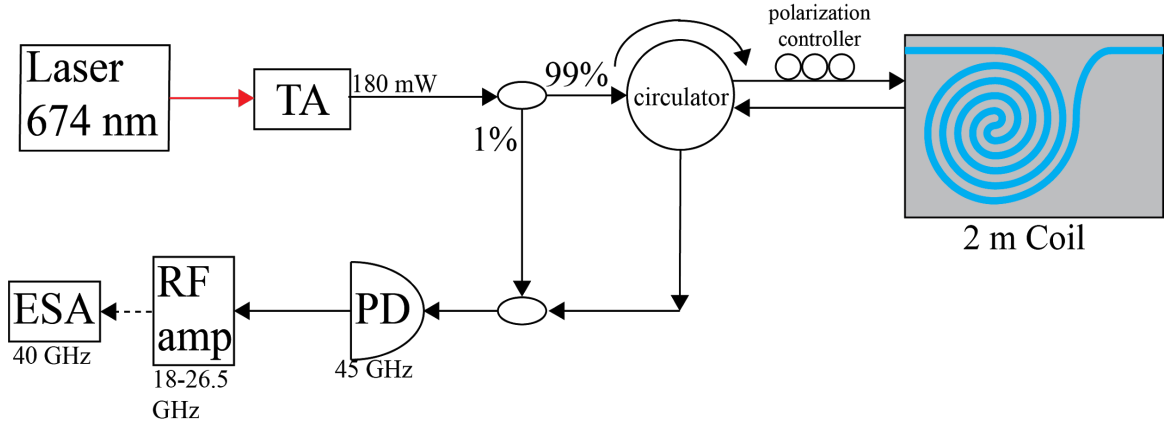

**Supplementary Fig. 1. Setup Schematic for Spontaneous Brillouin scattering spectrum.** Setup for measuring spontaneous Brillouin scattering. The 674 nm laser power is amplified with a tapered amplifier (TA), 1% of power is tapped out to beat with the Stokes tone which is obtained from the circulator. A high bandwidth 45 GHz photodiode (PD) is used to detect the 25.11 GHz stokes tone and is amplified using a large 40 dB gain radio frequency amplifier onto a 40 GHz electric spectrum analyzer (ESA).

## Supplementary Note 3: SBS lasing measurement setup

The measurements of Brillouin gain spectrum before and after threshold are made with the same setup as for spontaneous Brillouin measurements with the differences being the laser being locked to SLS cavity and the Rf amp is not used after PD. This was done to avoid pump noise from affecting heterodyne beat note (convolution of pump and Stokes) on the ESA. This is the only measurement in which the pump was locked.

For the frequency noise measurements, we use an optical frequency discriminator (OFD) and the details are described in the Methods section. The laser is not locked to any stable cavity for FN measurements and is not Pound-Drever-Hall (PDH) locked to the resonator; therefore, the noise suppression comes from the SBS phase noise suppression only. The device was on a temperature stage with 0.1 mK resolution to tune to and stabilize the resonance (Vescent Slice QT).

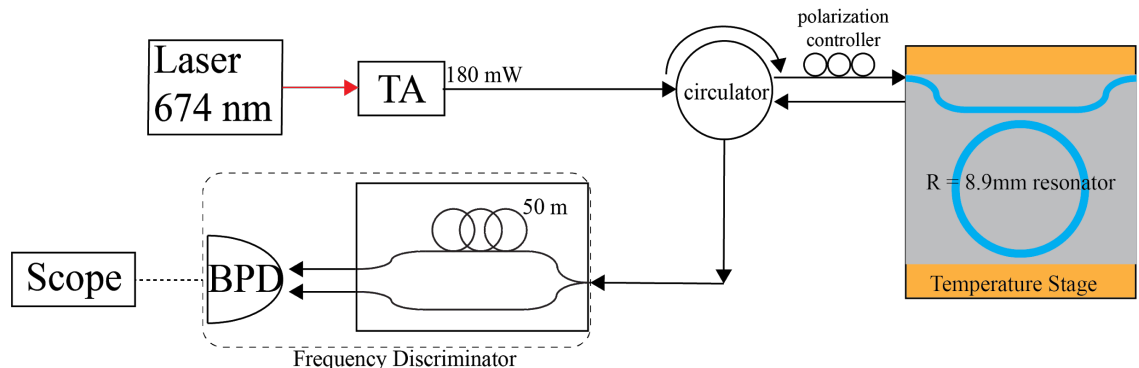

**Supplementary Fig. 2. Setup Schematic for SBS lasing stokes power and FN measurement.** Setup for measuring frequency noise (FN). Details of frequency discriminator used for measurement in methods section.

**698 nm clock SBS laser measurements.** The apparatus for the initial demonstration of a photonic integrated SBS laser at 698 nm is shown in Fig. 3. A continuous-wave Ti:sapphire laser is tuned near the clock wavelength. At 698.4 nm, the maximum output power is approximately 700 mW. Pump light is coupled through an optical fiber to the input of the bus waveguide. Light polarization is adjusted in the fiber to transverse-electric (TE) for maximum coupling to the resonator. As shown in Fig. 3, backward-propagating light is directed to an OSA and an ESA through the use of a quarter waveplate and polarizing beam splitter. The light propagating through the bus waveguide to the other side of the photonic chip is measured with a photodetector to monitor the resonance condition. The OSA is a scanning Fabry-Perot spectrometer (Bristol Instruments) that functions as a wavemeter and spectrum analyzer with ~5 GHz resolution. A photodetector (Thorlabs, 25 GHz 3dB bandwidth) converts the SBS laser light to an electrical signal through a beat note with the pump as shown in Fig. 3. The temperature of the photonic chip is maintained at +/- 0.01 °C by using a thermo-electric cooler.

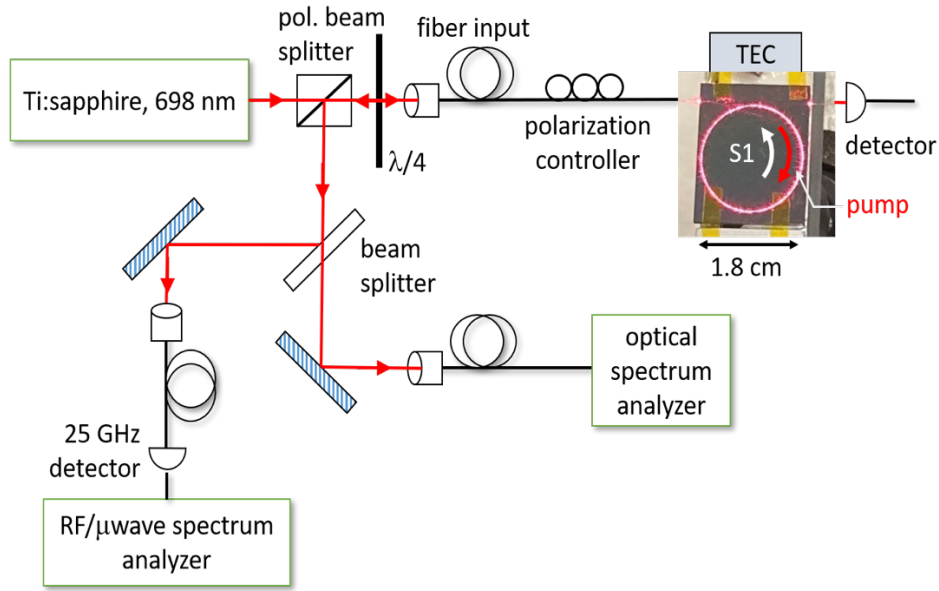

**Supplementary Fig. 3. Setup Schematic for SBS lasing stokes power measurement.** Setup for measuring Stokes power and Brillouin frequency shift in resonator at 698 nm. A thermo electric coupler (TEC) is used to tune and stabilize the resonance.

## Supplementary Note 4: Waveguide mode profiles at 674 nm and 698 nm

The waveguides are designed to support a single transverse electric (TE) mode in low loss, high aspect ratio $Si_3N_4$ waveguide[1,2]. The waveguide cross section and mode profiles of the TE mode is plotted in Fig. 4. The same waveguide geometry shown in Fig. 4a supports a TE mode at both 674 nm and 698 nm wavelengths. The bend loss is negligible for our resonator designs. The mode areas in our dilute low index contrast waveguides is 3.0 μm at 674 nm and 3.3 μm at 698 nm. This means that same waveguide supports SBS operation at both wavelengths by just changing ring radius and coupling gaps. Same core height can also be used for different wavelengths but changing the width to support single TE mode. The width, ring radius and coupling gap are all mask level changes.

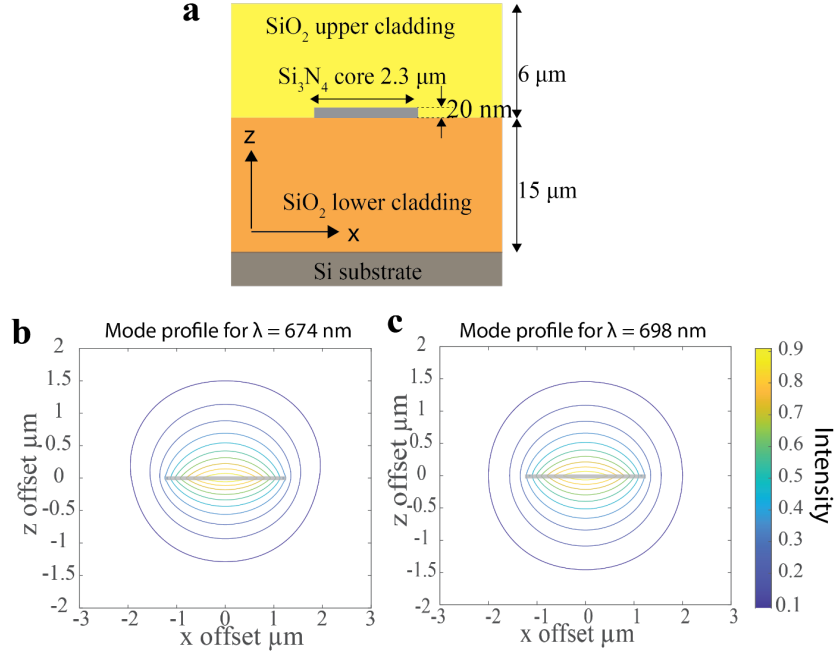

**Supplementary Fig. 4. Waveguide cross section and mode profiles at 674 nm and 698 nm. a,** Waveguide cross section with dimensions. **b,** Contour plot of simulated mode profile at 674 nm, waveguide represented in grey at center. **c,** Contour plot of simulated mode profile at 674 nm, waveguide represented in grey at center.

## Supplementary Note 5: SBS lasing stokes power modelling

We use the simplified coupled mode approach to model the intracavity photon number dynamics and the equations of motion for the optical field amplitude are given by[3]

$$\dot{a}_m = -(\frac{\gamma}{2} + \mu \mid a_{m+1} \mid^2 - \mu \mid a_{m-1} \mid^2)a_m + \sqrt{\gamma_c}F_{pump}\delta_{m0},$$ (1)

where the total loss rate, $\gamma = \gamma_{in} + \gamma_c$, is the sum of the intrinsic and coupling loss rates, μ is the Brillouin amplification rate per pump photon, and $F_{pump} = \sqrt{P_{pump}/h\nu}$ is the photon influx field amplitude. The Brillouin amplification rate μ of the SBS resonator is a function of Brillouin gain factor and resonator parameters,

$$\mu = \frac{h\nu v_g^2 G_B}{2L},$$ (2)

where $G_B = g_B/A_{eff}$ is the Brillouin gain factor, $g_B$ is the bulk Brillouin gain and $L$ is the resonator roundtrip length. From Eq. (1), we can find the threshold power for S1,

$$P_{th} = \frac{h\nu\gamma^3}{8\mu\gamma_c},$$ (3)

and the threshold power for S2 and S3 are $4P_{th}$ and $8P_{th}$.

The fundamental linewidth for S1 before its clamping is found to be,

$$\delta\nu_1 = \frac{n_0\gamma}{4\pi \mid a_1 \mid^2},$$

(4)

and it reaches its minimum at the S1 clamping point,

$$\delta\nu_1(4P_{th}) = \frac{n_0\mu}{2\pi}.$$

(5)

where $n_0 = k_m T/\hbar\Omega$ is the thermal occupation numbers of the acoustic mode. $n_0$ for the acoustic frequency 25.110 GHz at room temperature is estimated to be ~1300. From the measured on-chip threshold power of 14.7 mW, the resonator intrinsic linewidth of 8.0 MHz, the resonator coupling linewidth of 8.0 MHz, the resonator mode area of 3.0 μm² and the ring radius of 8.95 mm, we calculate that the estimated Brillouin gain rate μ is 50.7 mHz, the estimated minimal fundamental linewidth of S1 is 10.5 Hz and the estimated Brillouin gain factor is 0.49 (Wm)⁻¹. The estimated $G_B$ of 0.49 (W m)⁻¹ is close to our simulated $G_B$ of 1.03 (W m)⁻¹ at the experimental Brillouin shift frequency of 25.036 GHz, showing a good agreement between experiment and physics simulation. The theoretical S1 threshold of 7.0 mW assuming our simulated $G_B$ of 1.03 (W m)⁻¹ is also close to the measured threshold of 14.7 mW.

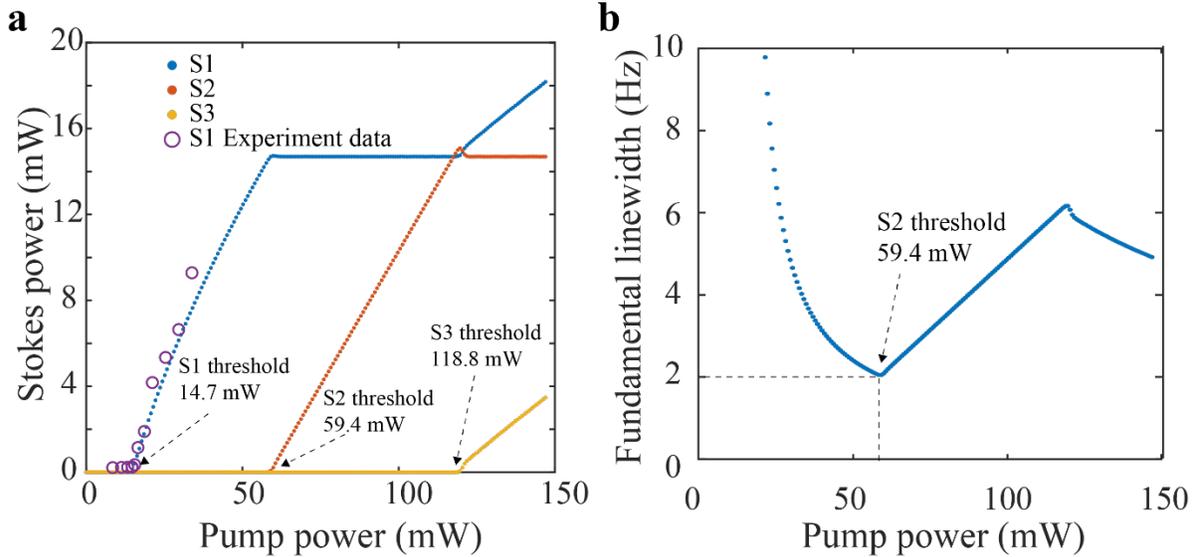

**Supplementary Fig. 5. Setup Schematic for SBS lasing stokes power measurement. a,** Modeled Stokes power for on chip power in range of 0-150 mW. **b,** Simulated fundamental linewidth of first stokes tone as a function of on-chip power assuming no crosstalk from pump. The fundamental linewidth decreases until S2 threshold is reached.